# UC2-ESP: A General-Purpose Framework for Open-Source Microscopy Control

**Benedict Diederich**[1,2], **Ingo Fuchs**[2], **Haoran Wang**[1,2], **Holger Bierhoff**[4,5], **Christian Kuttke**[2], **Rainer Heintzmann**[1,3]

Affiliations:
1. Leibniz Institute of Photonic Technology, Albert-Einstein-Str. 9, 07745 Jena, Germany
2. openUC2 GmbH, Hans-Knöll-Str. 6, 07745 Jena, Germany
3. Institute of Physical Chemistry and Abbe Center of Photonics, Friedrich-Schiller-University Jena, Helmholtzweg 4, 07743, Jena, Germany
4. Institute of Biochemistry and Biophysics, Center for Molecular Biomedicine (CMB), Friedrich Schiller University, Hans-Knöll-Str. 2, 07745, Jena, Germany
5. Leibniz Institute on Aging, Fritz Lipmann Institute, Beutenbergstr. 11, 07745, Jena, Germany

## Abstract

Building the optical setup for investigating biological questions comes with challenges. A major such challenge is setting up and synchronizing the control of multiple hardware components such as stages, cameras and lasers. With UC2-ESP we present a compact electronics system powered by the ESP32 microcontroller, designed to provide general-purpose control for various components in microscopy setups. Our system can interface with stepper motors, directed current (DC) motors, lasers (transistor-transistor logics, TTL or pulse width modulation, PWM), light emitting diodes (LEDs), and analog voltage outputs (galvo mirrors, led current control), allowing precise control over microscopy hardware. The platform is highly flexible, supporting custom pin configurations and multiple communication interfaces such as Bluetooth, universal serial bus (USB-serial), and HTTP via a built-in Webserver. A PlayStation controller can be used for haptic hardware manipulation, while commands are transmitted in a human-readable JSON format to ensure modularity and extensibility. The firmware is designed to receive parameters and execute actions dynamically, supporting complex control loops such as motor homing, stage scanning and temperature regulation via integrated controllers. Furthermore, the system integrates seamlessly with ImSwitch as well as MicroManager and offers a browser-based control tool using Web Serial. This open-source firmware enables microscopy research groups to develop custom setups and expand functionality efficiently, at low cost and high flexibility.

## The need for control - in microscopy

Modern microscopes have remained conceptually similar for more than a century, aiming to provide insights at the microscopic scale for biology, life sciences, and materials science. Yet present-day instruments are no longer just optical devices: they increasingly resemble compact, application-specific robots. Advanced systems contain motorised XY and Z stages, piezo drives, galvanometric mirrors, adaptive optics, heaters, cameras, and multiple illumination sources ranging from LEDs to pulsed lasers. All these subsystems must interact at millisecond precision, much like the coordinated motion control in 3D printers. The ability to perform accurate time-lapse experiments, high-speed imaging, or super-resolution methods requires tight synchronization between illumination, detection, and mechanical movement[1–5].

This complexity places microscope developers in a role akin to "full-stack developers": they must combine optical design, mechanical engineering, electronics integration, low-level firmware development, and high-level software for experimental workflows[6]. The final hurdle is often the orchestrated digital control of all components, for example, moving a stage while simultaneously triggering a laser and synchronising camera exposure to minimise photobleaching. Commercial controllers and DAQ cards exist, but they are expensive, often closed source, and rarely integrate well with modern workflows that rely on Python-based analysis, machine learning, or real-time feedback.

The maker community has shown that mass produced, off-the-shelf hardware can be repurposed: budget 3D printers provide precise motorised rails, and by replacing the extruder with a camera or attaching entirely new tool heads, platforms like the EnderScope[7], HistoEnder[8], or the Opentrons-contained microscope[9] demonstrate how one can built task specific scientific instruments at a fraction of the cost of bespoke instruments. These printers rely on established G-code standards and mature motion-planning firmware such as GRBL, Marlin, or Klipper. However, while highly optimised for extrusion and motion control, these large codebases are difficult to adapt for microscope-specific components like multiple light sources, filter wheels, or sensitive detectors. Unlike 3D printing, microscopy lacks a universal control language: every stage, laser, camera, or piezo typically comes with its own proprietary driver or dynamic linked library (DLL), often locked behind closed USB protocols. Therefore, scientists who build custom microscopes must be familiar with a variety of device-specific libraries and programming patterns in order to ultimately obtain microscopic images or create more complex workflows. Custom prototypes complicate matters further: one setup may only require a TTL pulse to gate a diode laser, while another light-sheet system may need four high-current stepper channels for sample scanning plus an additional axis for automated focusing. Off-the-shelf controllers rarely expose "hackable" interfaces, and discontinued vendor hardware often ends up as electronic waste.

These issues, combined with the necessity to digitise our own modular UC2 microscopes, motivated the design of UC2-ESP: a modular, hardware agnostic, fully open-source firmware and electronics framework. Our approach is guided by seven principles:

1. cross-platform operation regardless of host OS, energy (e.g. battery driven)
2. stand-alone capable hardware-control
3. broad support for off-the-shelf components common in microscopy hardware,
4. deployment on inexpensive, readily available microcontrollers,
5. installation-free debugging requiring only a web browser,
6. a friction-less setup procedure that works out of the box, and
7. straightforward hooks for diverse experimental workflows and protocols.

By combining configurable hardware modules (stepper drivers, PWM laser controllers, temperature regulators) with the ESP32 microcontroller, UC2-ESP[10] consolidates drivers for lasers, stages, heaters, sensors, and auxiliary opto-electronics into a single lightweight binary.

Devices are exposed via Wi-Fi, Bluetooth, CAN, I²C, or USB under a unified JSON/REST command layer. Instead of needing one controller or port per device, the system reduces the hardware footprint to a single interface, while remaining extensible and hackable. This allows both novice and expert users to rapidly build, modify, or repurpose experimental setups and integrate them seamlessly with environments such as Python or GUI platforms like ImSwitch[11].

# Methods

## System Architecture

We chose the ESP32 as the central element for driving various electronic components - such as stepper motors, LED arrays (e.g., for phase-contrast microscopy), TTL/PWM signals for laser control, or sensor evaluation via I²C and distributed operation via CAN bus. This compact microcontroller comes with a wide range of interfaces (UART, WiFi, Bluetooth, I²C, CAN bus, etc.) and benefits from a large open-source development community in which many code examples and drivers are available[12]. Since the ESP32 lacks the necessary power electronics to directly drive high-current components such as stepper motors or lasers, additional hardware is required. We therefore provide two complementary solutions. The first leverages standardized adapter boards widely used in CNC systems and 3D printers - most notably the CNC Shield v3 (Protoneer, New Zealand) combined with an ESP32-WEMOS D1 (various manufacturers, China). To accommodate different hardware configurations, pin assignments are decoupled from the main firmware, allowing flexible customization. The second solution is a series of custom-developed extension boards (i.e. UC2e), designed specifically for microscopy setups. This board integrates support for different communication protocols and is optimized for the combination of actuators, light sources, and sensors typically required in modular (UC2) microscopes as shown in Figure 1.

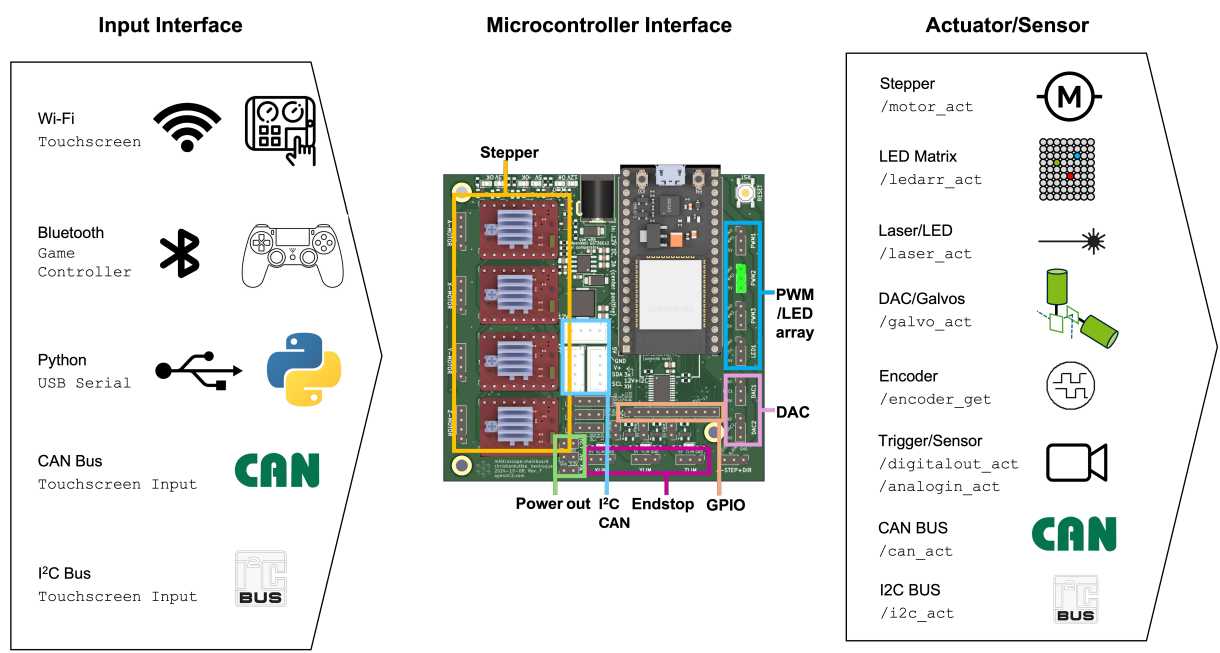

**Figure 1:** Input-Output Diagram. The microcontroller accepts various input interface(left) and can drive different actuators and sensors(right).

UC2-ESP distinguishes time-critical hardware synchronisation, which is handled on the microcontroller itself, from user-directed parameter updates that can tolerate network latency. The latter are delivered asynchronously over Wi-Fi, USB-Serial, or Bluetooth. For example, a PlayStation 4 joystick can adjust laser power in real time via Bluetooth, a smartphone can send the same command through the ESP32's access-point mode, and a Python script can stream updates over a wired serial connection. All messages use a self-descriptive JSON syntax and a REST-like endpoint scheme (e.g., `/motor_act`), making the API both human-readable and trivially extensible. Although JSON adds a modest payload overhead, it greatly simplifies debugging and future expansion while keeping the integration burden on client software to an absolute minimum (Figure 2).

The firmware is built using PlatformIO[13], an open-source community-based framework to write and compile firmware for different microcontrollers13, providing a reproducible development environment and simplifying the integration of both the ESP-IDF toolchain12 and Arduino components14. The programming consists of several modules, a structure that follows the classic Arduino approach with a setup() function (initialization) and a loop() function for each such module. The globally running programming loop runs through individual module loops consecutively. This allows multiple tasks to run concurrently - for example, when using a closed-loop control mechanism for stepper motors. Each module's setup() routine is executed at initialization.

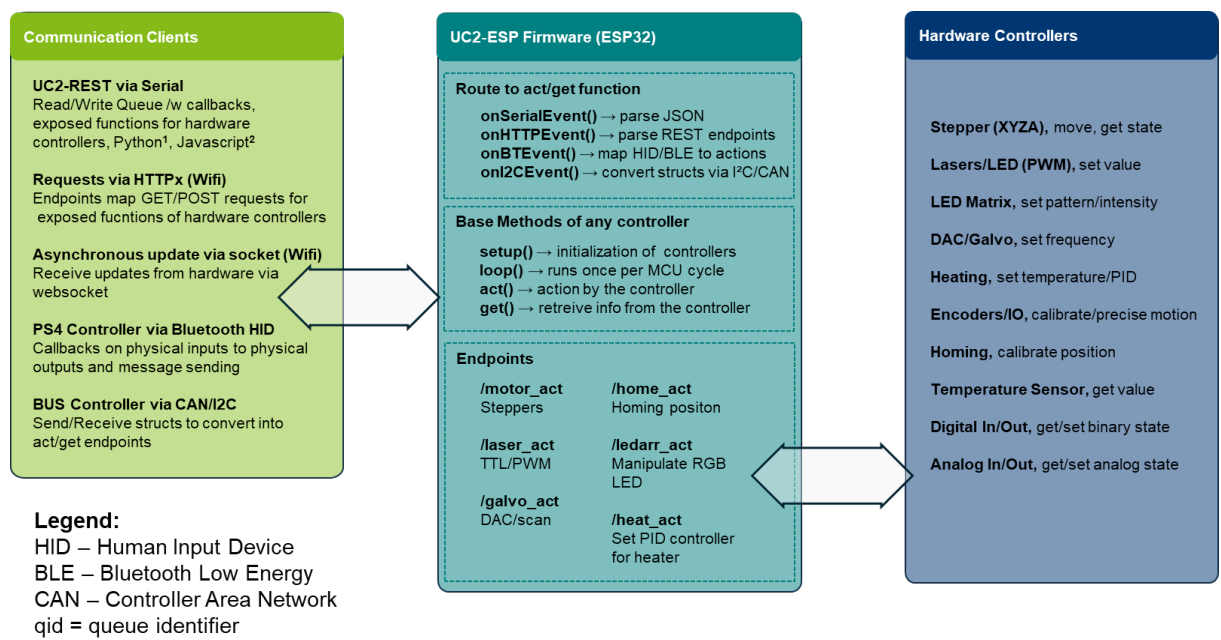

**Figure 2:** Firmware structure. The firmware supports multiple communication clients and decodes task functions into executable actions for the hardware.

To keep the memory footprint small and tailor the configuration to various applications, modules and communication interfaces are switched on or off using preprocessor directives (`#define`). WiFi, Bluetooth and USB-Serial can thus run simultaneously or be used exclusively to optimize performance. Certain tasks can also be offloaded to satellite boards connected via I²C or CAN bus - for instance, handling complex stepper motor control loops, including acceleration profiles and time-critical closed-loop regulation. In such cases, the master board (ESP32) merely handles the communication, while the specialized secondary board carries out the control tasks. This modular principle conserves memory and computing resources on the main system while still enabling efficient implementation of sophisticated control requirements.

Each request sent from the host to the ESP32 firmware includes a unique request ID (qid) also visualized in Figure 3. Upon receiving a well-formed JSON request, the ESP32 immediately returns an interim response echoing the qid, asynchronously before executing the command. This acknowledgement ensures that malformed or lost requests can be identified e.g. in case a request cannot be parsed (e.g. invalid JSON), the ESP32 responds with a negative qid to indicate an error.

Once the request has been fully executed, the ESP32 issues a final response associated with the same qid. For example, if the host requests a motor to move to a new position, the interim acknowledgement confirms receipt of the request, and the final response confirms that the motor has reached its target (or that the request was superseded by a newer one).

In addition to responding to host-initiated requests, the ESP32 can also send asynchronous updates to the host, such as reporting a new motor position when moved by an external input (e.g. joystick). On the host side, these unsolicited messages can be bound to callback functions, allowing the system state to remain synchronized without constant polling. In effect, the communication model combines a traditional request/response pattern with event-driven updates, giving both sides the ability to exchange messages as needed.

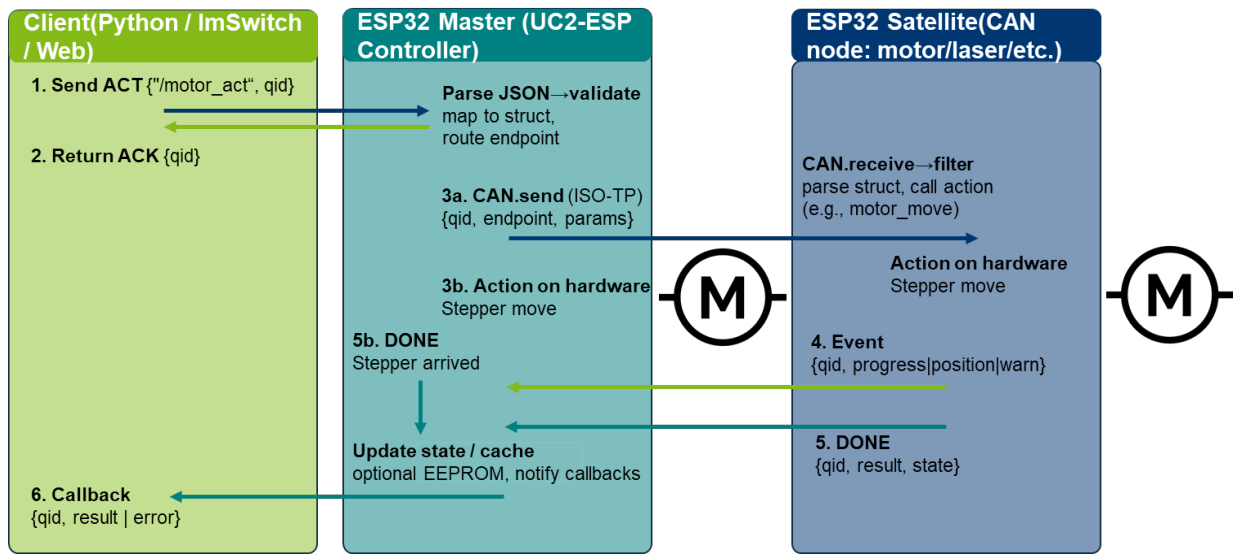

**Figure 3:** Task flow of the firmware. An operation initiated by the client generates an "ACT" message, which is transmitted to the firmware to request execution of a function. The firmware parses the received JSON string, extracts the task, and translates it into a hardware-specific action. This command is then delivered to the corresponding hardware module for execution. Upon successful completion, the firmware transmits a response message back to the client, thereby confirming task execution.

## Firmware Features

### REST-Like Interface

The firmware's API is inspired by common REST practices, offering a unified, platform-independent interface for controlling actuators and reading sensors. Two types of commands exist: an "ACT" (analogous to a POST) that triggers a function, and a "GET" (similar to a GET request) that retrieves current system states. An example "ACT" command to move two stepper motors simultaneously by a given number of steps might look like:

```
{
  "task": "/motor_act",
  "motor": {
    "steppers": [
      {"stepperid": 1, "position": 10000, "speed": 5000, "isabs": 0, "isaccel": 0},
      {"stepperid": 3, "position": 10000, "speed": 5000, "isabs": 1, "isaccel": 0}
    ]
  }
}
```

In this case, motor 1 moves 10,000 steps at 5,000 steps/second in a non-accelerated, relative motion, while motor 3 moves to position 10,000 with the same parameters. The firmware parses the JSON string, starts the motors in a non-blocking manner, and immediately returns a success message. A second message is sent once the final positions are reached, including updated positions. An optional queue ID (`qid`) can be included to correlate commands and responses.

A "GET" request, such as

```
{"task": "/motor_get", "qid":1}
```

retrieves the current state of each motor (e.g., enabled status, position):

```
{
        "motor":
                {"steppers":[
                        {"stepperid":0,"position":0,"trigOff":0,"trigPer":-1,"trigPin":-1,
                "isStop":0,"isRunning":0,"isDualAxisZ":0,"isforever":0,"isen":0,"stopped":1},
                        {"stepperid":1,"position":188990,"trigOff":0,"trigPer":-1,"trigPin":-
                1,"isStop":0,"isRunning":0,"isDualAxisZ":0,"isforever":1,"isen":0,"stopped":1},
                        {"stepperid":2,"position":117315,"trigOff":0,"trigPer":-1,"trigPin":-
                1,"isStop":0,"isRunning":0,"isDualAxisZ":0,"isforever":1,"isen":0,"stopped":1},
                        {"stepperid":3,"position":-70926,"trigOff":0,"trigPer":-1,"trigPin":-
                1,"isStop":0,"isRunning":0,"isDualAxisZ":0,"isforever":1,"isen":0,"stopped":1}]
                        },"qid":1
}
```

Similar requests can quickly poll sensor data, making it straightforward to integrate real-time monitoring of temperature, light intensity, or other parameters. This way, host software can work asynchronously by relying on events sent by the microcontroller or synchronously, by actively polling the microcontroller's state.

**Connecting multiple Controllers on the Microcontroller**
Within the firmware, different modules can be linked to enable automated workflows. For instance, temperature control[15], where the reading from a digital temperature sensor (e.g., DS28xx) is fed into a PID controller to regulate a PWM heater, maintaining a stable environment (e.g., for a stage-top incubator). Similarly, a home-position routine connects a digital input controller (endstop) to a stepper motor controller, instructing the stepper to move until the endstop is triggered and then reset its position to zero. The resulting controller provides a new REST-like endpoint (inspired by lab-things[16]), while the execution and computation of control actions (e.g. in case of the temperature control) is fully carried out on the microcontroller, hence saving resources when sending/receiving control commands. An exemplary homing command might look like:

```
{
  "task": "/home_act",
  "home": {
    "steppers": [
      {"stepperid": 2, "timeout": 20000, "speed": 15000, "direction": -1, "endposrelease": 3000}
    ]
  }
}
```

This instructs the firmware to move stepper 2 at 15,000 steps/second in the specified direction, check for the endstop, and then backs off by 3,000 steps once triggere, demonstrating how real-time control routines are directly executable on the microcontroller unit (MCU) level through the same REST-like interface.

**Integration of Satellites via I²C**

To support more complex workflows and reduce the computational overload on the main controller, UC2-ESP firmware allows external devices to be networked over I²C. Since I²C is a standard interface for sensors and actuators, integrating new peripherals such as a temperature or pressure sensor becomes straightforward. Developers can simply attach the device on the I²C bus and access it via the existing JSON/serial interface without major firmware changes.

Beyond simple peripherals, UC2-ESP also supports the creation of custom I²C "satellites". These can be secondary microcontrollers such as another ESP32 configured as I²C slaves with their own device address. A typical use case is offloading real-time, closed-loop tasks such as detailed stepper acceleration profiles to a satellite board, while the primary ESP32 remains responsible for high-level coordination and communication (WiFi, Bluetooth, USB). This division of labor reduces both memory and processing load on the primary MCU, while still exposing the satellite's functionality (e.g. motor control) through the same unified JSON/I²C interface. In this way, UC2-ESP makes it equally simple to integrate commodity I²C devices or to extend the system with custom controllers that interoperate seamlessly with the rest of the firmware.

**Integration of Satellites via CAN**

Unlike I²C, where the master must actively poll each slave for status updates, a Controller Area Network (CAN) supports true bidirectional messaging: any node can transmit, and all nodes on the bus receive the frame but only act on messages addressed to them. Widely adopted in the automotive industry, CAN uses a differential signal pair that is tolerant of electrical noise and higher supply voltages, ideal for setups where 12 V motor drivers share cabling with low-level logic. The ESP32 includes a native CAN (TWAI) peripheral, and a transceiver module is used to translate between logic and bus levels.

In UC2-ESP we assign CAN identifiers (IDs) to each class of component (e.g. master node: 0x00, motors: 0x10-0x19, lasers: 0x20-0x29). Every node implements both transmit and receive handlers. Upon receiving a message, each device filters the frame headers and processes only messages with IDs relevant to its function, thereby avoiding unnecessary parsing of unrelated traffic. The design follows a request/response pattern for most interactions: the "master" (e.g. a controller interfacing with Python or ImSwitch) sends requests to a satellite (e.g. move motor X to position N), and the addressed device responds when the request is acknowledged or completed. However, satellites can also issue messages on their own, for instance, a motor controller broadcasting that it has reached its target position, or a sensor publishing a new measurement without being explicitly polled. This mix of directed requests and asynchronous state updates reduces bus traffic while keeping all participants synchronized.

Because the JSON-based UC2-ESP protocol often exceeds the 8-byte payload of a single CAN frame, we employ ISO-TP[17] segmentation to split large commands across multiple frames, which the receiver then reassembles into a single action. Sending and receiving tasks run concurrently in separate queues, enabling reliable handling of simultaneous traffic. Broadcasts (messages intended for all nodes) are supported through a reserved ID. Figure 3 illustrates a mixed UC2-ESP network where motors, sensors, and lasers share the same two-wire bus, coordinated through ID assignment.

**Integration with Software**

From the outset, the firmware has been designed to be as user-friendly and straightforward as possible, minimizing additional software requirements. Because the ESP32 supports various communication methods (HTTP requests, USB Serial, Bluetooth), users can interact with the device in multiple ways:

1. **Serial Interface**
   The simplest and most robust method involves a wired USB-Serial connection, allowing direct communication between the host software (e.g., Python, ImSwitch, MicroManager, Javascript) and the microcontroller. We provide a Python-based library (UC2-REST[18]) that wraps each endpoint or device into a dedicated function call, which then generates the required JSON string. For example, a function call in Python might look like `move_stepper(stepper_id=1, distance=1000)`, which internally constructs the JSON command and sends it to the firmware. This library also supports optional callbacks for events such as target position reached or updated sensor readings.
2. **Browser-Based WebSerial**
   To avoid any installation overhead, we have developed an online control tool that runs in a standard web browser and communicates with an ESP32 Master using WebSerial (Youseetoo.github.io). This interface provides a simple GUI for controlling motors, LEDs, and lasers without needing to install drivers or additional software. WebSerial allows direct communication with the ESP32's USB-Serial port, so end users can open a web page, connect to the device, and start issuing commands in real time. Through GitHub Actions, the firmware can be compiled automatically and hosted on a static (github) webpage, making new releases instantly accessible.The ESP32 Flashing tool, based on work from the home assistant community[19], makes use of the newly established Web Serial Standard and helps the user to install firmware on their board variant.
3. **Integration with ImSwitch**
   The firmware integrates seamlessly with ImSwitch, enabling advanced scripting and on-the-fly control of microscope components. Through Python, users can quickly set up automated imaging routines, such as moving a stage to multiple locations or adjusting laser intensity in response to live image feedback.
4. **Integration with Micro-Manager**
   Similar to the integration with ImSwitch using Python, a dedicated C++ device driver has been developed to interface with Micro-Manager in a manner akin to a "hub device." Lasers, stages, and additional sensors can be controlled via serial commands passed directly between Micro-Manager and the ESP32. This design makes it easy to integrate UC2-ESP hardware into existing microscopy workflows without extensive software modifications.

5. **Integration as a Web-request based control**
   The ESP32 is well known for its ability to be used in Internet of Things applications (IoT) due to its internal Wifi Module. We implement a simple web server that exposes the different endpoints that control the hardware as http-request endpoints. The ESP32 can act as an access point or connect to an available WiFi network. Additionally, a static webpage shipped via this web server renders buttons controlling the hardware. This is very helpful for automated control of a microscope, with the smartphone camera acting as the acquisition device.
6. **Bluetooth Controller-based control**
   In addition to Wifi, the ESP32 also offers the use of Bluetooth. For this, we implemented a library that can connect to various Human Interfacing Devices (HID), where buttons and joysticks (e.g. from the PS4 controller) are then mapped to hardware functions. This realizes an intuitive control of the microscope when for example searching the sample area for a specific feature by steering the joystick.

Overall, these integrations ensure that users can operate the firmware in environments ranging from lightweight browser-based setups to comprehensive imaging platforms, all while relying on the same unified JSON-based control structure and the same codebase.

**Time-critical processing of hardware control commands in a sequence**

Although UC2-ESP already permits millisecond-level coordination of motors, lasers, and cameras, it does not yet expose a generic, user-defined "timeline engine" that would let a client upload arbitrary synchronisation table with conditional branches or time-outs. Time-critical routines are instead hard-coded in firmware: for the galvo scanner, for example, a dedicated `for` loop driven by a hardware timer toggles SPI-controlled DAC voltages at a fixed pixel dwell time, while the XY stage scanner moves through a list of coordinates and emits a camera-trigger pulse at each stop. These blocks run outside the main task, rely on the ESP32's deterministic timer modules, and may employ lightweight ISRs (interrupt service routines) to flip pins precisely without jeopardising the RTOS scheduler. The user can populate such blocks at run-time by passing a JSON structure e.g.

```
{
  "task": "/motor_act",
  "stagescan": {
    "coordinates": [{"x":100,"y":200}, {"x":300,"y":400}],
    "tPre":50, "tPost":50,
    "illumination":[50,75,100], "led":100
  }
}
```

or record a linear macro of several commands and replay it multiple times:

```
{
  "tasks":[

{"task":"/motor_act","motor":{"steppers":[{"stepperid":1,"position":1000,"speed":20000,"isaccel":1,"accel":500000}]}},
    {"task":"/laser_act","LASERid":3,"LASERval":200},
    {"task":"/laser_act","LASERid":3,"LASERval":0},
    {"task":"/state_act","delay":1000}
  ],
  "nTimes":2
}
```

During execution these routines are blocking: the serial interface is muted until completion to guarantee timing accuracy. In practice this approach delivers micro- to millisecond synchrony for common tasks (galvo line scans, grid acquisitions, stage-top autofocus), but extending it to fully programmable, conditional timelines remains future work. Planned improvements include a lightweight on-board scheduler that would accept event tables with absolute or relative timestamps, timeout logic, and branching conditions which would enable complex experiment choreography without recompiling firmware while still leveraging hardware timers and ISR hooks for sub-millisecond determinism. This is more suitable for FPGA-based approaches in microFGPA[20].

## Use case

The UC2-ESP firmware has already been deployed in a range of open-hardware and commercial instruments, highlighting its versatility across very different imaging modalities and actuator-sensor constellations. Below we name a few use cases, where the firmware was used under different platform-io configurations (noted in the paragraphs).

**Structured-Illumination Module on a Commercial Stand (openSIMMO) - DIY and off-the-shelf automation**
Wang et al.[21] presented an open-source structured-illumination extension depicted in Figure 4a driven entirely by a ESP32 Wemos D1 board in combination with the CNC Shield v3 board running UC2-ESP *[env:UC2_WEMOS]*. Two fiber-coupled diode lasers (488 nm and 638 nm) receive TTL modulation from the board, while a NEMA-17 stepper motor that was coupled to the microscope's fine-focus knob translates the objective for axial focus stacking. A repurposed 3D-printer heater bed keeps the sample chamber at 37 °C, its temperature held by an on-board PID loop with a temperature sensor. The same firmware was fully integrated into the automated SIM imaging workflow, where a dedicated controller was responsible to trigger time-lapse sequences and 3D stacks[22]. Key actuators/sensors: 2 × TTL-gated lasers, 1 × Z-axis stepper, 1 × thermistor/heater pair. PIO Firmware Config: `[env:UC2_WEMOS]`

**Reviving Scrap-Heap Microscopes**

The Genomic Vision FiberVision (France) microscope (Fib. 4b) became unusable after vendor software support was discontinued, despite its intact mechanical and optical subsystems. UC2-ESP was used to regain control of the motorized XYZ stage (Physik Instrumente, Germany) through standard STEP/DIR stepper interfaces and a customized adapter cable to bridge the UC2 ESP32 board with their stepper motors. The original fluorescence light engine (Lumencor, Spectra X 6s, USA) was re-integrated via its RS-232 protocol bridged by the ESP32 UART, while additional extensions such as objective turret drivers and filter revolvers (SmarAct, Germany) were added over their proprietary USB interface. Once linked to ImSwitch, the revived instrument could perform automated routines including multi-plane Z-stacks and tiled acquisitions. This case demonstrates how UC2-ESP enables the sustainable repurposing of discontinued commercial systems into flexible, Python-controlled imaging platforms, thereby extending hardware lifetimes and reducing e-waste. PIO Firmware Config: `[env:UC2_3]`

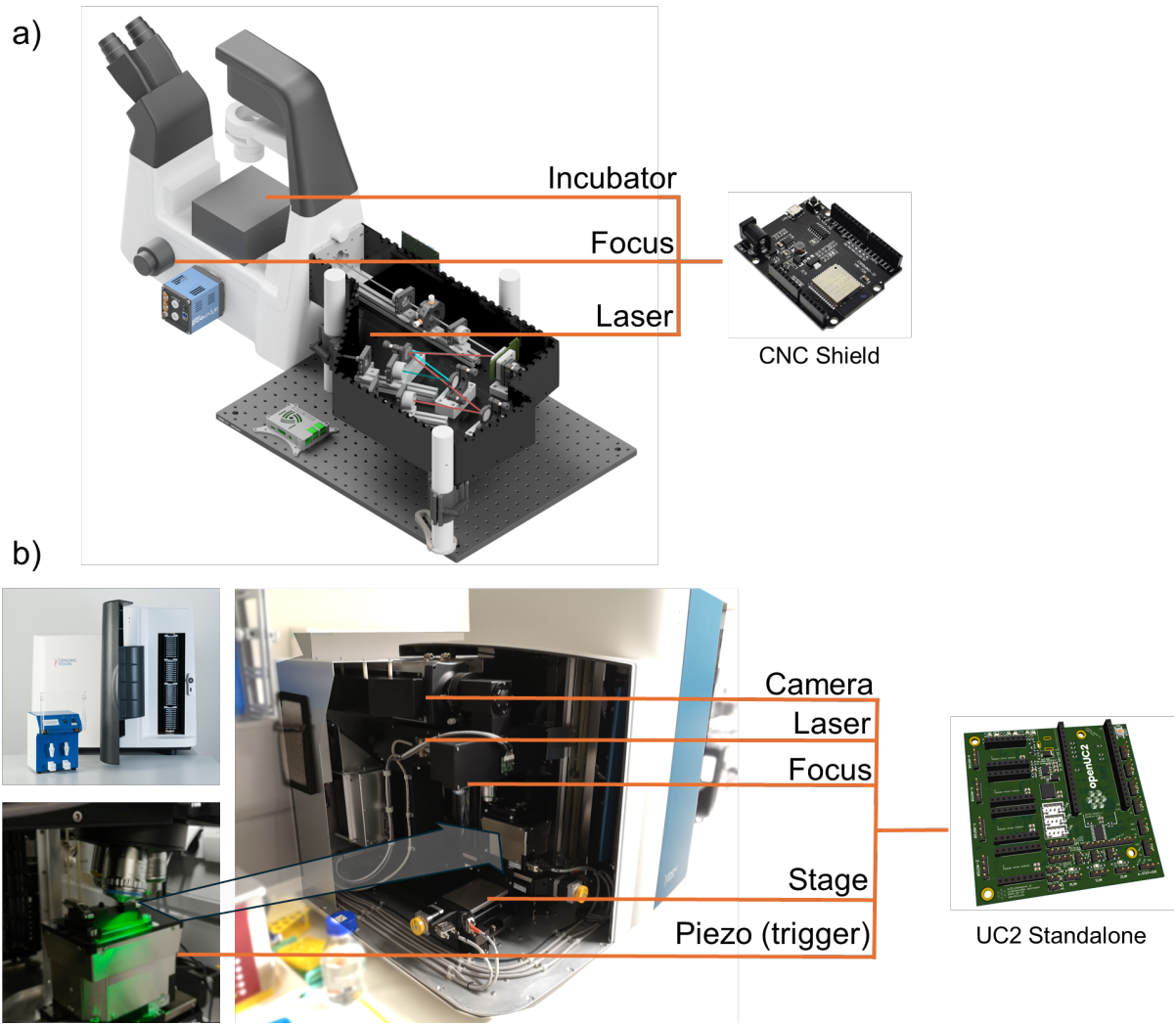

Figure 4 Use cases with customized microscopes. a) An open-source structured illumination microscopy extension controlled by the firmware, b) a revived multi colour fluorescence microscope for automated slide scanning.

**Cost-Effective Single-Molecule Localization & Modular Light-Sheet Microscopy**

Zehrer et al.[23] combined high-numerical-aperture objectives with low-cost lasers and opto-mechanics to create a single-colour super-resolution system using STORM (stochastic optical reconstruction microscopy) capable of single-particle tracking (Figure 5a). A low-cost high-power laser diode and a white LED are switched consecutively via ESP-generated TTL signals to acquire dSTORM and brightfield illumination images; three TMC-2209 stepper channels drive the XY translation stage and Z focus. This is again orchestrated in ImSwitc[11], a Python interface that grants full access to the functions of the board. Alternatively, the same setup can be controlled from micromanager with the corresponding device driver. Key actuators/sensors: 1 × TTL-laser, 1 × white-LED, 3 × steppers with end-stops, PIO Firmware Config: [env:UC2_WEMOS]

Multiple labs pair UC2-ESP with ImSwitch for customised light-sheet setups constructed from UC2 cubes as in Figure 5b. A motorised XYZ stage carries the specimen through a static laser sheet whose intensity is TTL-modulated; another stepper adjusts detection focus. A ring of NeoPixels provides bright-field illumination for coarse alignment. The acquisition of the Z-stack is carried out such that the beginning and the end of the motion is synchronized with the frame number, so that individual frames are mapped along the stack equidistantly allowing the software to map every camera frame to an exact Z coordinate. Joystick control handled by the built-in Bluetooth stack lets users quickly centre the region of interest before handing control back to an automated Python routine. The on-board digital-analog converter (DAC) can provide fast oscillating scanning signals for galvo mirrors to generate a scanning-based light-sheet. Key actuators/sensors: 1 × TTL-laser, 1 × LED, 4 × steppers, PIO Firmware Config: `[env:UC2_3]`

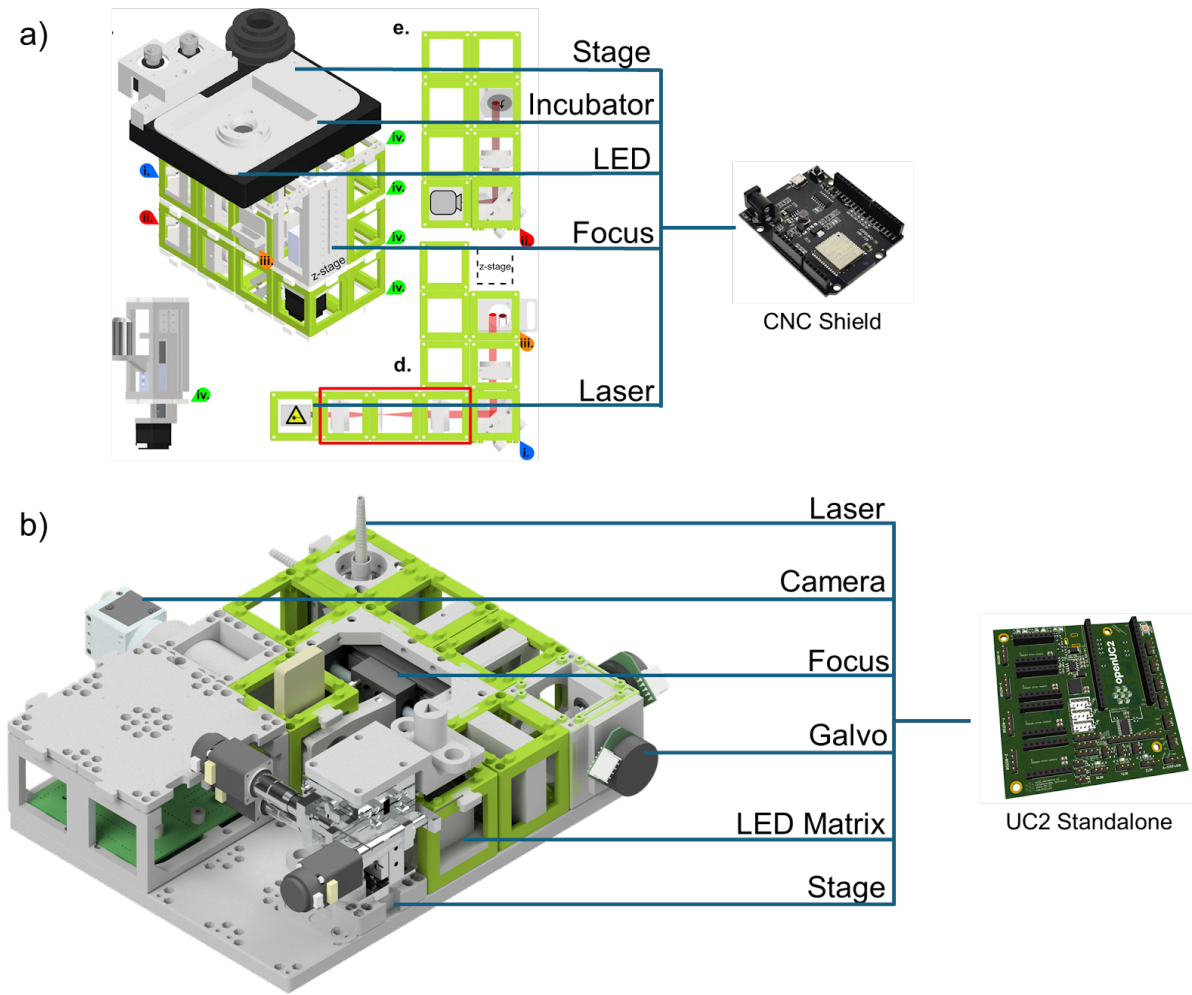

Figure 5 Use cases with UC2 cube-based fluorescence microscopy systems. a) dSTORM microscope, b) light-sheet microscope.

**Rapid Prototyping for Start-Ups and Educational Kits**

Several early-stage companies adopted UC2-ESP to shorten their prototyping cycle. openUC2 (Jena, Germany) contributed with a firmware update that extended the CAN implementation with the ISO-TP-enabled CAN interface, so a single twisted pair now carries commands to satellite boards controlling stage-top incubators, motors, or NeoPixel illumination rings (Figure 6). Because every firmware flavour shares the same JSON/REST API, hardware variants can be swapped without changing higher-level code or GUIs. Seeed Studio (Shenzhen, China) likewise ships an educational microscope that leverages the same core to drive auxiliary pumps and RGB arrays for classroom demonstrations. Off-loading closed-loop temperature or pump control to tiny ESP32S3 (Xiao) CAN satellites frees the main ESP32 for high-bandwidth USB and Wi-Fi tasks, markedly improving overall stability. PIO Firmware Config: `[env:UC2_3_CAN_HAT_Master, env:seeed_xiao_esp32c3_can_slave_motor]`

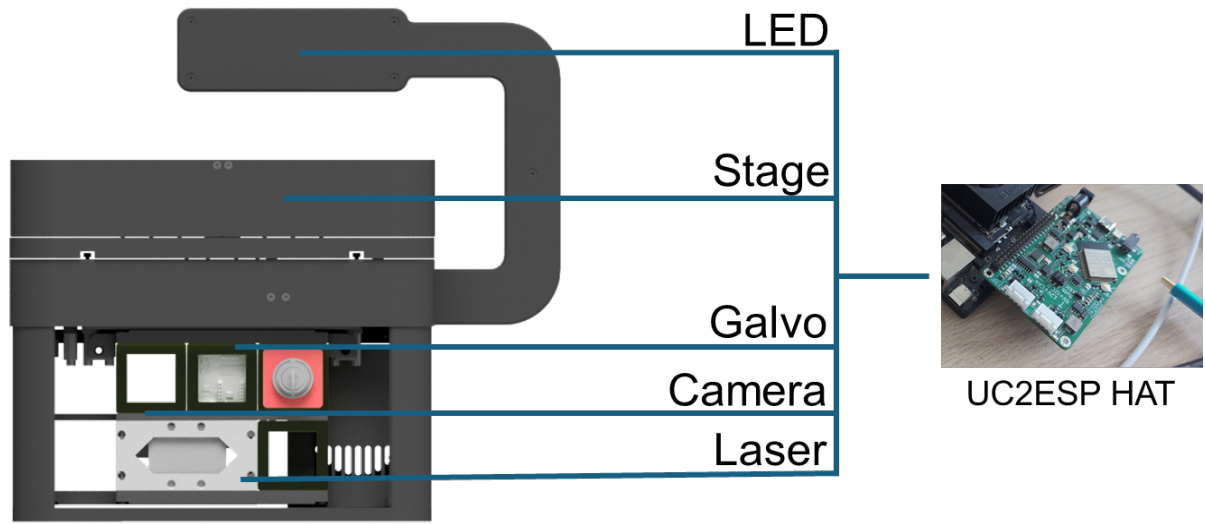

Figure 6 An automated microscope relies on the UC2-ESP firmware. UC2ESP HAT is employed as master and each hardware has its own slave ESP for task communication.

## Conclusion

We believe this firmware fills a gap between inexpensive open-source microscopes and the growing community focused on "smart" microscopy techniques. Rather than prioritizing minimum latency, we emphasize modularity, enabling researchers to mix and match different actuators, sensors, and communication interfaces. In addition to supporting professional experiments such as temperature regulation or advanced scanning routines, our system is also well-suited for teaching and demonstration purposes, offering accessible ways to learn about automated microscopy.

Future development will focus on expanding bus-based communication (e.g., I²C, CAN) to incorporate additional specialized hardware components such as spatial light modulator (SLM). By distributing tasks across multiple boards, we can further improve real-time performance and reduce the computational burden on the main MCU. We anticipate that this flexible, open-source approach to firmware development will continue to stimulate innovation in the broader microscopy community, allowing rapid prototyping and iterative improvement of custom experimental setups.

# Supplementary

## Legend — UC2-ESP Bus Layers

- **GPIO** = native MCU pins (STEP/DIR, PWM, digital I/O)
- **I²C / SPI** = onboard or remote peripheral link
- **CAN** = remote module over CAN-FD network
- **Analog** = voltage or current output from DAC

## Table of available components

| Component | Category | Key operations / functions | Representative commands | Interface |
|---|---|---|---|---|
| **Stepper motor (X / Y / Z / A)** | Actuator | relative / absolute move continuous isforever run stop, enable/disable driver setpos, softlimits, dual-axis sync high-level stagescan sweeps | /motor_act, motor_get /home_act (homing) | Native STEP/DIR GPIO I²C (remote driver) CAN (remote module) |
| **Linear-encoder** | Sensor | read position PID assisted moveP for feedback-looped positioning calibration calibration of step/distance relation, setup | /linearencoder_get, /linearencoder_act | A/B, I²C |
| **Incremental encoder** | Sensor | read position zero / calibrate use calliper-based absolute positioning | /encoder_get /encoder_act | GPIO |
| **Home / end-stop** | Sensor | initiate homing move set polarity, timeout | /home_act /home_get | GPIO (motor/digitalin) |

| Device | Type | Functions | Endpoints | Interface |
|---|---|---|---|---|
| **Laser driver (0-3)** | Actuator | analogue power LASERval anti-speckle modulation (pwm variation, amplitude/period) PWM/servo freq set | /laser_act /laser_get | PWM GPIO / DAC |
| **LED array (WS2812 ring / matrix / strip)** | Actuator | individual RGB, rings, halves, circles, off global intensity | /ledarr_act /ledarr_get | GPIO (1-wire) |
| **Digital output (trigger)** | Actuator | assign pin edge-timed trigger (DelayOn/Off) reset trigger | /digitalout_set, /digitalout_act | GPIO |
| **Digital input** | Sensor | read pin state | /digitalin_get | GPIO |
| **DAC channel** | Actuator | set fixed value waveform (frequency,offset,amplitude, shape: sine, rectangle, triangle) | /dac_act /dac_act_fct | $I^2C$ / SPI |
| **Galvo scanner** | Actuator | raster scan (min/max, step, dwell time, frames) | /galvo_act /galvo_get | Analog ±10 V via DAC |
| **Heating control** | Actuator + Sensor | PID on/off, set Kp/Ki/Kd, target °C read temperature ds18b20 temperature sensor | /heat_act /heat_get | GPIO (PWM) + lasercontroller |
| **Objective changer** | Actuator | calibrate (dir, polarity) toggle, explicit move set slots x1/x2, focus z1/z2 | /objective_act, /objective_get | Stepper GPIO (motor+digitalin) |
| **Rotator stage** | Actuator | multi-axis rotate moves use 28byj-48 motors | /rotator_act /rotator_get | Stepper GPIO |
| **TMC driver (per axis)** | Actuator config | micro-step, RMS current stall-guard, cool-step params control via UART | /tmc_act /tmc_get | UART |
| **CAN-bus node** | Comm. | set / query node address restart, motor proxy cmds handle communication via ISO-TP | /can_act /can_get | CAN (TWAI) |
| **$I^2C$-bus scan** | Comm. | enumerate attached devices send messages read messages | /i2c_get | $I^2C$ |
| **Parallel bus** | Comm. | data transfer | | |
| **System state** | Sensor | heap, busy, debug, restart, delay get build information | /state_get /state_act | — |

| Focus-scan routine (motor + LED + trigger) | Composite scan | axial sweep Z-start → nZ steps<br>timed pre/trigger/post delays<br>LED illumination mask | /motor_act -> focusscan | Stepper GPIO + One-wire GPIO |
| Stage-scan routine (motor + laser + LED + trigger) | Composite scan | raster XY grid<br>per-pixel trigger, LED or laser patterns<br>high-speed coordinated motion | /motor_act -> stagescan | Stepper GPIO + Analog (DAC) + One-wire GPIO |

## Data availability

| Description | Link |
| --- | --- |
| **Firmware Repository (UC2-ESP)** | https://github.com/youseetoo/uc2-esp32 |
| **Online-based Firmware Flashing** | https://youseetoo.github.io/ |
| **Online-based Firmware Testing** | https://youseetoo.github.io/indexWebSerialTest.html |
| **Python interface for UC2-ESP Firmware (UC2-REST)** | https://github.com/openUC2/UC2-REST |
| **Configuration Files for different board configurations; Configuration header and platformio.ini file** | e.g. https://github.com/youseetoo/uc2-esp32/blob/2e8c58d836c9aa9ce97831c19465f411cf8b873e/main/config/UC2_3_CAN_HAT_Master/PinConfig.h<br><br>And https://github.com/youseetoo/uc2-esp32/blob/2e8c58d836c9aa9ce97831c19465f411cf8b873e/platformio.ini#L654 |

## Competing Interests

B.D. is co-founder of the company openUC2 that builds and distributes open-source microscopes. Both B.D. and H.W. are shareholders of that company. All other authors declare no competing interests.